\documentclass[11pt]{article}

\usepackage[T1]{fontenc}
\usepackage[utf8]{inputenc}
\usepackage[a4paper,margin=1in]{geometry}
\usepackage{newtxtext,newtxmath}
\usepackage{microtype}
\usepackage{amsmath,bm}
\usepackage{graphicx}
\usepackage{float}
\usepackage{authblk}
\usepackage[colorlinks=true,linkcolor=blue,citecolor=blue,urlcolor=blue]{hyperref}
\usepackage{cleveref}
\usepackage{cite}
\usepackage{siunitx}
\usepackage{pdfpages}

\newcommand{\ket}[1]{\left|#1\right\rangle}

\newcommand{\expval}[1]{\left\langle #1\right\rangle}
\newcommand{\matrixel}[3]{\left\langle #1 \middle| #2 \middle| #3 \right\rangle}

\title{Aziz and Howl's Gravity-Induced Entanglement Channel Is Essentially Classical Mechanics}

\author[1,5]{Hanyu Xue\thanks{These authors contributed equally.}\thanks{Corresponding author: \texttt{xhy2002@mit.edu}}}
\author[2,3,4]{Ziqian Tang$^*$}
\author[5]{Chen Yang}
\author[6]{Zizhao Han}
\author[7]{Zikuan Kan}
\author[2]{Yulong Liu\thanks{Corresponding author: \texttt{liuyl@baqis.ac.cn}}}

\affil[1]{Department of Physics, Massachusetts Institute of Technology, Cambridge, MA 02139, USA}
\affil[2]{Beijing Key Laboratory of Fault-Tolerant Quantum Computing, Beijing Academy of Quantum Information Sciences, Beijing 100193, China}
\affil[3]{Beijing National Laboratory for Condensed Matter Physics, Institute of Physics, Chinese Academy of Sciences, Beijing 100190, China}
\affil[4]{University of Chinese Academy of Sciences, Beijing 100049, China}
\affil[5]{School of Physics, Peking University, Beijing 100871, China}
\affil[6]{Center for Quantum Information, IIIS, Tsinghua University, Beijing 100084, China}
\affil[7]{School of Physics, Renmin University of China, Beijing 100872, China}

\date{April 16, 2026}

\begin{document}

\maketitle

\begin{abstract}
Aziz and Howl argued that a classical gravitational field can generate quantum entanglement through a quantum-field-theoretic channel mediated by virtual matter propagation.  However, their claimed channel is more naturally and accurately understood as semiclassical wavepacket motion in an external gravitational field, rather than as a distinctively quantum-field-theoretic entangling effect. Moreover, the result of their perturbative computation is incorrectly magnified:
they selected a discontinuous wavefunction with infinite kinetic energy as the initial state and simultaneously treated it as stationary. Once a correct treatment using Gaussian wavepacket is adapted, the resulting effect will be negligibly small.
\end{abstract}

\section{Introduction}

\begin{figure}[htbp]
    \centering
    \includegraphics[width=0.82\linewidth]{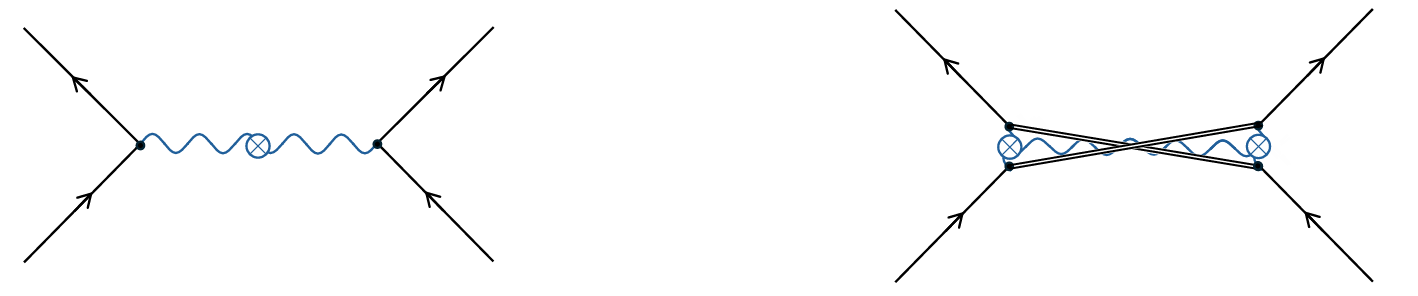}
    \caption{Direct and exchange diagrams discussed in Ref.~\cite{aziz_classical_2025}. The external arrowed lines denote incoming and outgoing matter particles, the wavy line denotes the classical gravitational field, and the internal double line denotes the matter propagator. Note that in a standard external-field expansion, these wavy lines represent classical insertions rather than a gravitational propagator; see Figure \ref{fig:external-insertions}.}
    \label{fig:exchange-diagrams}
\end{figure}

Aziz and Howl argued in \cite{aziz_classical_2025} that, within quantum field theory, a classical gravitational field can generate quantum entanglement between two macroscopic objects. In their picture, the relevant mechanism is a quantum communication channel mediated by virtual matter propagators. More specifically, they consider the two Feynman diagrams shown in Figure \ref{fig:exchange-diagrams}: the external arrowed lines represent incoming and outgoing matter particles, the internal double line denotes the propagator of the matter field, and the wavy line denotes the classical gravitational field. In their classical-gravity treatment, $h_{\mu\nu}$ enters as a classical external field rather than a quantum dynamical field, so there is no Wick contraction of the gravitational field and hence no gravitational propagator. Their claim is that, because the particles are identical, the two propagation topologies are indistinguishable, and the crossed diagram yields entanglement.

In \cite{tang2025mattermediatedentanglementclassicalgravity}, we argued that this interpretation is fundamentally mistaken. Since the process involves neither particle creation nor annihilation, the relevant dynamics are already those of ordinary quantum mechanics. We further pointed out that modeling the matter field as a free scalar field before the gravitational perturbation is unphysical in the mechanical settings relevant to gravity experiments. Such experiments typically involve solids, whose constituent particles are confined by binding potentials rather than described by a free field. Once the binding potential is included, the process identified in \cite{aziz_classical_2025} becomes a tunnelling process whose amplitude is exponentially suppressed with the separation between the objects and becomes appreciable only at atomic length scales; see also Appendix~\ref{appendixTunnelling}.

In an unpublished reply, Aziz and Howl later clarified that their model was not intended to describe solids, but rather systems without binding potentials, such as photons, cold atomic gases, and Bose--Einstein condensates.%; see also Appendix~\ref{appendixReply}.
That clarification makes it worthwhile to re-examine the physical content of their argument in the unbound case. We show below that their result rests on a mathematically inconsistent and physically unmotivated choice of initial state: they adopt a wavefunction with infinite kinetic energy and then treat it as quasi-stationary. Once this inconsistency is removed, the purported effect collapses to ordinary motion under gravity, for which classical mechanics provides a non-perturbative and substantially more accurate description than the isolated perturbative diagram highlighted in \cite{aziz_classical_2025}.

Our argument has three steps. Section 2.1 restates their entanglement criterion, Section 2.2 gives a classical upper bound on the crossed channel, and Section 2.3 shows that their power-law result arises from an inconsistent initial state together with an unjustified stationary approximation.

\section{Main Argument}

\subsection{Entanglement criterion in Aziz and Howl's setup}

To make the discussion transparent, we first restate Aziz and Howl's criterion for entanglement. In \cite{aziz_classical_2025}, each object contains $N$ elementary particles. For present purposes, however, each object may be replaced by a single particle without changing the physical mechanism under discussion. 
They assume that the first particle has left and right localized modes $1L$
and 
$1R$, while $2L$
 and 
$2R$ denote those of the second particle; the closest pair is $1R$ and $2L$, separated by $d$.
The initial state is assumed to be the product state
\begin{equation}
    |\psi_0\rangle=(\ket{1L}+\ket{1R})\otimes(\ket{2L}+\ket{2R})
\end{equation}
containing four branches. 
The corresponding amplitudes $\alpha_{LL}$, $\alpha_{LR}$, $\alpha_{RL}$, and $\alpha_{RR}$ are propagation amplitudes obtained by evolving the initial wavepackets by $U(t)$ and projecting them onto the same initial localized modes\footnote{See Eq. (13) in their Supplementary Material of \cite{aziz_classical_2025}.}:
\begin{equation}
    \alpha_{ij}=\langle 1i|\otimes\langle2j| U(t)|1i\rangle\otimes|2j\rangle
\end{equation}

 Aziz and Howl interpret entanglement through these four amplitudes, such as a non-negligible determinant $\alpha_{LL}\alpha_{RR}-\alpha_{RL}\alpha_{LR}$.

Because the two particles are assumed to be identical, the amplitude for each particle to remain in its original location is indistinguishable from the crossed amplitude in which the two particles propagate into one another's modes---for example, from $1R$ to $2L$. In their interpretation, a nonzero crossed contribution to $\alpha_{ij}$ is therefore a necessary condition for entanglement. In diagrammatic language, this is the contribution of the exchange diagram in Figure \ref{fig:exchange-diagrams}, with the external lines integrated over the localized modes. In wavepacket language the same point is more transparent: one particle's wavepacket must evolve so as to overlap the initial localized mode of the other particle. If the whole wavepackets move far away from their initial positions, then the original mode basis is no longer a useful way to characterize entanglement.%\footnote{Also see the second footnote of their reply, Appendix \ref{appendixReply}.}. 
Although possible, we do not attempt to define entanglement in that case in order to avoid potential debates. Nonetheless, for the present discussion, it is enough to note that Aziz and Howl's cross-propagation mechanism requires a non-negligible overlap between the two wavepackets, otherwise direct and exchange Feynman diagrams can be clearly distinguished. The question is therefore when such overlap can become appreciable.

\subsection{Classical estimate of the crossed channel}

In a slowly varying external field, the evolution of a wavepacket has two components: spreading due to the uncertainty principle, and motion of the packet center driven by the external force. Aziz and Howl effectively neglect the former, while the latter is already captured by ordinary classical dynamics. We focus on the amplitude $\alpha_{RL}$: since the distance $d$ between local modes $1R$ and $2L$ are shortest, the corresponding effect is strongest. Let the radii of the two wavepackets be $R\ll d$. Denoting by $g\sim GM/d^2$ the typical gravitational acceleration between $1R$ and $2L$, the packet centers are displaced by
\[
\delta x\sim \frac{1}{2}gt^2.
\]
Using the parameters quoted by Aziz and Howl, $t=2\,\mathrm{s}$, $d=200\,\mu\mathrm{m}$, and $M=10^{-14}\,\mathrm{kg}$, one finds
\[
\delta x\sim 10^{-13}d.
\]
The crossed propagation required by their mechanism is therefore negligible by any reasonable experimental standard in the parameter regime of interest.

The same conclusion can be formulated within standard quantum field theory \cite{zielinski-2024,sterman-1993,Rajeev_2021,hattori-2023}. In an external classical field, the propagator of the matter field $\phi$ is dressed by an infinite series of interaction insertions, while the gravitational field itself has no propagator. Resumming these insertions yields the dressed propagator familiar from the Furry/Schwinger proper-time treatment, which describes the semi-classical motion of the wavepacket. The right-hand diagram in Figure \ref{fig:exchange-diagrams} is only one fourth-order term in this perturbative expansion.

\begin{figure}[htbp]
    \centering
    \includegraphics[width=0.78\linewidth]{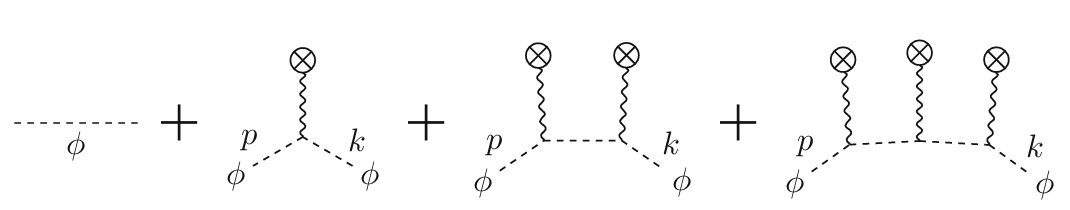}
    \caption{External-field insertions on a matter propagator, reproduced from Eq.~(5) of Ref.~\cite{zielinski-2024}.}
    \label{fig:external-insertions}
\end{figure}

Yet Eq.~(70) of the Supplementary Material of \cite{aziz_classical_2025} produces a power-law dependence on time, in direct conflict with the semi-classical expectation that the displacement should remain negligible when $\delta x\ll d/2$. This discrepancy is not a minor artifact of truncating the perturbation series. It reflects two basic errors in the treatment of the initial state. 

\subsection{Two errors in the perturbative treatment}

The first error lies in the choice of initial state itself. When Aziz and Howl assume that the matter is distributed within a sphere of radius $R$ around the origin, they take the initial wavefunction to be the step function
\begin{equation}
    \psi(\mathbf{r})=\frac{1}{\sqrt{V}}\theta(R-r).
\end{equation}

Step function may serve as a good approximation for potentials, but not for wavefunctions. This wavefunction is discontinuous, lies outside the domain of the Hamiltonian, and has an infinite expectation value of the kinetic energy. From Eq.~(22) of the Supplementary Material of \cite{aziz_classical_2025}, its Fourier transform behaves asymptotically as
\begin{equation}
\tilde{\phi}(\mathbf{k})\sim \frac{\cos kR}{k^2}. \label{eq:phi-tail}
\end{equation}
Aziz and Howl describe this as a rapidly decaying function and on that basis adopt a nonrelativistic approximation. That description is untenable. The decay is not rapid enough: not only does $\expval{p^2}$ diverge, even $\expval{p}$ diverges:
\begin{equation}
\expval{p}=\int_0^\infty \hbar k\,|\tilde{\phi}(\mathbf{k})|^2\,4\pi k^2\,dk\sim \int_0^\infty \frac{dk}{k}=\infty. \label{eq:p-diverge}
\end{equation}

For fixed $t$ and $r\to\infty$, the asymptotic form of the real-space wavefunction $\psi(r,t)$ is semi-classically governed by $\tilde{\phi}(k)$ evaluated at momentum $\hbar k=mr/t$. More precisely, methods such as the Imaging Theorem  give\cite{Feagin_2017,del-campo-2009}
\begin{equation}
|\psi(\mathbf{r},t)|^2\sim \left(\frac{m}{i\hbar t}\right)^3\left|\tilde{\psi}\!\left(\frac{mr}{\hbar t}\right)\right|^2\sim \frac{t}{r^4}. \label{eq:imaging}
\end{equation}
Hence
\begin{equation}
\expval{r}=\int_0^\infty r\,|\psi(r,t)|^2\,4\pi r^2\,dr\sim \int_0^\infty \frac{dr}{r}=\infty. \label{eq:r-diverge}
\end{equation}
In other words, this initial state immediately develops tails extending to arbitrarily large distances and cannot consistently be treated as quasi-stationary.

For a quasi-stationary wavepacket, one should instead choose a Gaussian initial state,
\begin{equation}
\psi(\mathbf{r})=\frac{1}{(2\pi R^2)^{3/4}}e^{-r^2/(4R^2)}, \label{eq:gaussian-initial}
\end{equation}
for which the probability density after a time $t$ is
\begin{equation}
|\psi(\mathbf{r},t)|^2=\frac{1}{(2\pi \sigma_t^2)^{3/2}}e^{-r^2/(2\sigma_t^2)}, \label{eq:gaussian-time}
\end{equation}
with
\begin{equation}
\sigma_t^2=R^2+\left(\frac{\hbar t}{2mR}\right)^2. \label{eq:sigma}
\end{equation}
This packet spreads asymptotically with speed $\hbar/(2mR)$ and retains Gaussian suppression outside the packet. The calculation in \cite{aziz_classical_2025} misses precisely this suppression because the chosen initial state already presupposes infinite kinetic energy.

The second error is between Eq.~(22) and Eq.~(23) of the Supplementary Material of \cite{aziz_classical_2025}, where they make the non-relativistic assumption $\hbar k\ll mc$. While this assumption is inconsistent with their infinite-energy state, the approximation they make is also inappropriate. They actually replace
\[
E=\sqrt{(\hbar k)^2c^2+m^2c^4}
\]
by $mc^2$, completely discarding the kinetic-energy term $\hbar^2k^2/(2m)$, and then replace the exact wavefunction (Supplementary Eq. (22)) by a stationary wavefunction (Supplementary Eq. (23)). However, that infinite kinetic term is precisely what controls wavepacket spreading. Once the stationary approximation fails, the further claim that the internal line must be interpreted as off-shell virtual-particle propagation has no basis: the crossed channel does contain the ordinary propagation of matter wavepackets.

\section{Conclusion}

Aziz and Howl proposed a quantum-field-theoretic mechanism by which a classical gravitational field could generate quantum entanglement. We explain this mysterious effect in a simpler and clearer way: once the initial state is treated correctly, the purported effect reduces to ordinary propagation under gravity. In this regime, classical mechanics is not merely an alternative description; it is the correct non-perturbative description, whereas the isolated perturbative diagram emphasized in \cite{aziz_classical_2025} is a distorted presentation of the same negligible physics.

For experiments that seek to infer the quantumness of gravity from entanglement, this channel must be excluded as a possible background. In practice, however, it is unlikely to generate an observable effect of its own. First, after replacing the unphysical step-function state by a Gaussian wavepacket and restoring the omitted kinetic term, one finds that the packet center accelerates in the gravitational field $g$ while the packet simultaneously spreads at speed $v=\hbar/(2mR)$. Under realistic parameters, both effects are extremely small, and the resulting contribution to entanglement is correspondingly negligible. Second, the crossed-propagation channel can be suppressed simply by placing a barrier between the two objects, which returns the problem to the bound-state and tunnelling picture discussed in \cite{tang2025mattermediatedentanglementclassicalgravity}. Third, once the overlap becomes appreciable, the two objects are already so close that direct contact and other uncontrolled interactions are likely to dominate the signal.

\section{Acknowledgments}
We acknowledge the support of National Natural Science Foundation of China (No.~12374325), Young Elite Scientists Sponsorship Program by CAST (Grant No.~2023QNRC001), and Beijing Municipal Science and Technology Commission (Grant No.~Z221100002722011). Z.~H. acknowledges the support of the National Natural Science Foundation of China (Grant No.~T2225008) and the Tsinghua University Dushi Program. No code or data were used or generated in this study.

\section{Data Availability}
This study involves no data that can be shared.

\section{Code Availability}
The main results do not rely on any generated code.

\bibliographystyle{plain}
\bibliography{apssamp}

\appendix

\section{Virtual propagation and Quantum Tunnelling}\label{appendixTunnelling}

Aziz and Howl suggested in their unpublished reply that we confuse quantum tunnelling and virtual (off-shell) propagation. Despite deviating a bit from our main text, we present a short note to clarify their relationship.

First, for ordinary solids or liquids treated as many-particle systems, binding potentials cannot be ignored, so one must consider tunnelling instead of free propagation. 

Second, although the definitions of off-shell propagation and tunnelling are not exactly the same, they indeed have close relationship, which is especially transparent in one-dimensional bound states.
Consider the one-dimensional Schr\"odinger equation, where $H_0$ is the free-particle Hamiltonian and $V$ is a binding potential that is nonzero only in a small region near the origin. In a form analogous to the Lippmann--Schwinger equation \cite{schwinger-1961}, we write
\begin{equation}
(H_0+V)\ket{\psi}=E\ket{\psi}. \label{eq:schrodinger}
\end{equation}
Its solution can be written as
\begin{equation}
\ket{\psi}=\frac{1}{E-H_0+i\epsilon}V\ket{\psi}. \label{eq:lippmann-schwinger}
\end{equation}
In the absence of the binding potential, the fixed-energy Green's function \cite{economou-2013} is
\begin{equation}
G_E(x-x')=\matrixel{x}{\frac{1}{E-H_0+i\epsilon}}{x'}. \label{eq:greens-function}
\end{equation}
Hence the bound-state wavefunction satisfies the exact relation
\begin{equation}
\psi(x)=\int G_E(x-x')V(x')\psi(x')\,dx. \label{eq:bound-state}
\end{equation}
Because $V(x)$ is nonzero only near the origin, the far-field wavefunction is approximately
\begin{equation}
\psi(x)\propto G_E(x). \label{eq:far-field}
\end{equation}
Writing the Green's function as a Fourier transform,
\begin{equation}
G_E(x)=\int_{-\infty}^{\infty}\frac{e^{ipx}}{E-p^2/(2m)+i\epsilon}\,dp, \label{eq:fourier-green}
\end{equation}
we see that for $E<0$ the propagator is off shell for all real $p$, so the integrand is smooth on the real axis. Its Fourier transform therefore decays as
\[
G_E(x)\sim e^{-\sqrt{2m|E|}\,|x|}\qquad (|x|\to\infty).
\]
In a residue evaluation, this comes from the poles at $p=\pm i\sqrt{2m|E|}$, which corresponds precisely to a semi-classical particle with imaginary momentum. By contrast, when $E>0$, on-shell values of $p$ exist on the real axis, and the poles on the real contour dominate the integral.

In this example, ``virtual propagation'' emphasizes that the momentum integral encounters no poles on the real axis, whereas ``quantum tunnelling'' emphasizes that, after contour deformation, the dominant contribution comes from poles with nonzero imaginary momentum. These are two descriptions of the same underlying physics. 

%\section{Aziz and Howl's previous reply}\label{appendixReply}

%For the reader's convenience, we attach the previous reply by Aziz and Howl mentioned in the main text. The following pages are reproduced verbatim for convenience and are not part of the present manuscript.

%\includepdf[pages={1-5}]{AHreply.pdf}

\end{document}